\documentclass[aps,prd,showpacs,floatfix,preprint,a4paper]{revtex4}
\usepackage{amsmath,bm}
\usepackage{graphicx}
\usepackage{epsfig}
\begin{document}

\title{Massive Neutron Stars with Antikaon Condensates in a Density Dependent Hadron Field Theory} 
\author{ Prasanta Char$^{\rm (a)}$, Sarmistha Banik$^{\rm (b)}$ }
\affiliation{$^{\rm (a)}$Astroparticle Physics \& Cosmology Division,
Saha Institute of Nuclear Physics, 1/AF Bidhannagar, Kolkata 700 064, India}
\affiliation{$^{\rm (b)}$BITS Pilani, Hyderabad Campus, Samirpet Mondal, Hyderabad 500078, India}

\begin{abstract}
The measurement of $1.97 \pm 0.04 M_{solar}$ for PSR J1614-2230 and $2.01 \pm 0.04M_{solar}$ for PSR J0348+0432 puts a strong constraint 
on the neutron star equation of state and its exotic composition at higher densities. In this paper, we investigate the 
possibility of exotic equation of state within the observational mass 
constraint of $2M_{solar}$ in the framework of relativistic mean 
field model with density-dependent couplings. We particularly study the effect of antikaon condensates 
in the presence of hyperons on the mass-radius relationship of the neutron star. 
\end{abstract}
\maketitle
\section{Introduction}
 
Neutron stars are fascinating objects to probe exotic states of dense matter 
that cannot be otherwise studied in a terrestrial laboratory. The central 
density of its core surpasses the nuclear density by a few times. Exact nature 
of its internal structure is yet to be understood. Various theoretical models have 
been proposed to explain its structure and characteristics. 
Among them Walecka model, a Lorentz covariant theory of
dense matter involving baryons and mesons, has been widely applied to study 
the neutron star matter \cite{Serot}. This traditional meson exchange picture is known as the relativistic field theoretical model. The model 
including non-linear scalar meson terms yields the saturation properties of nuclear matter and finite nuclei quite well. 
However, regime above saturation density is not well understood. Extrapolating the nuclear matter properties to high density leads 
to uncertainties. In most of the relativistic mean field (RMF) calculations, 
non-linear self interaction terms for scalar and vector fields are introduced 
to account for the high density behaviour \cite{Sch}. But this may not be a reliable 
approach due to instabilities 
and higher order field dependence that may appear at high densities. Another 
more suitable approach is to incorporate the density-dependence through the meson-baryon 
couplings \cite{ddrh,typel05,typel09}. In the density dependent model the 
appearance of a rearrangement term in baryon chemical potential significantly 
changes the pressure, consequently the equation of state (EoS) at higher densities. 

We must also consider the role of nuclear symmetry energy, the energy associated with the isospin asymmetry, on the behaviour of the EoS at high densities.
The nuclear symmetry energy alters the stiffness of the EoS. 
It is of great importance, along with its density dependence, in studying many crucial problems in astrophysics, such as neutronization
in core collapse supernova explosion, neutrino emission from protoneutron star (PNS), neutron star radii, crust thickness, cooling among various others 
\cite{Lim}. The symmetry energy and its density dependence near the saturation density $n_{0}$ are denoted by $S_{\nu}=E_{sym}(n_{0})$ 
and slope parameter $L=3n_{0}dE_{sym}/dn|_{n=n_{0},T=0}$ respectively. These parameters can be constrained by the findings of precise 
nuclear physics experiments (heavy ion collision analysis, 
dipole polarizability analysis etc.) as well as  astrophysical 
observations. The bounds on the parameters are found to be 
29 MeV $< S_{\nu} <$ 32.7 MeV and 40.5 MeV $< L <$ 61.9 MeV  
respectively \cite{Lim,hempel}.
Now if we look into the most popular and widely used parametrizations to model neutron star structure, such as GM1, TM1, NL3 etc., we find that 
the values of both symmetry energy and its slope parameters in all these 
cases (For GM1, $S_{\nu}=32.47$ MeV and $L=93.8$ MeV; 
TM1, $S_{\nu}=36.95$ MeV and $L=110.99$ MeV; NL3, $S_{\nu}=37.39$ MeV and $L=118.49$ MeV \cite{hempel}) do not quite fall into the 
experimental range. Whereas the density dependent (DD2) RMF model, we are 
going to employ in this paper with $S_{\nu}=31.67$ MeV 
and $L=55.04$ MeV, are fully consistent with the above experimental and observational constraints \cite{typel09}. In fact, it is the 
only relativistic EoS model with linear couplings. Also the DD2 EoS model agrees well with the predictions by Chiral EFT \cite{hempel}.
However it should be noted that the density dependent parametrization (DD) was in 
use \cite{Fuc95,Len95,ddrh}
even before this symmetry energy experimental data was available. The current DD2 model differs from the previous DD model only 
by the use of experimental nuclear masses \cite{typel09}.

The discovery of  binary pulsar PSR 1913+16 in 1974 by Hulse and Taylor lead to first precise measurement of neutron star mass 
($1.4408 \pm 0.0003 M_{solar}$) \cite{hulse}. The millisecond pulsar PSR 1903+0327 of $1.67 \pm 0.02M_{solar}$ \cite{champ}, measured in 2008, PSR J1614-2230 
of $1.97 \pm 0.04 M_{solar}$ \cite{demo} in 2010 and PSR J0348+0432 of $2.01 \pm 0.04M_{solar}$ \cite{anton} subsequently in 2011 have raised the bar. 
The knowledge of the precisely measured mass of neutron stars has important consequences for constraining the equation of state of
dense matter. It can throw light on otherwise poorly known composition of the compact star core. 

It is still an open issue if novel phases of matter such as hyperons, Bose-Einstein condensates of pions and kaons and also quarks may exist 
in neutron star interior or not. The presence of hyperons and antikaon 
condensates makes the EoS softer resulting in a smaller maximum mass neutron 
star than that 
of the nuclear EoS \cite {Gle,third}. In fact strangeness in the high-density baryonic matter is almost the inevitable consequence of 
Pauli principle. Strange degrees of freedom would be crucial for long time evolution of the PNS \cite{sn_meta} also. The observation 
of massive compact stars with mass $>2M_{solar}$ puts stringent constraint on the model of neutron stars and may abandon most of 
the soft EoS. However, it is at present not possible to rule out all exotica with recent observation as many model calculations 
including hyperons and/or quark matter could still be compatible with the observations. Many of these approaches are parameter 
dependent, for example the EoS with hyperons are compatible with the benchmark of $2M_{solar}$ \cite{taur,oertel,Weiss,Last,Armen}.
Antikaon condensate is another possible strange candidate in the dense interior of neutron stars. It was first demonstrated 
by Kaplan and Nelson within a chiral $SU(3)_L \times SU(3)_R$ model in dense matter formed in heavy ion collisions \cite{Kap}. 
The isospin doublet for kaons is $K \equiv (K^+,K^0)$ and that for antikaons $\bar K \equiv (K^-, {\bar K}^0)$. The attractive 
interaction in nuclear matter reduces the in-medium energy of (anti)kaons; at higher density eventually falls below 
the chemical potential of the leptons and replace them. Antikaon condensation was later studied in details in the context of 
cold neutron star and protoneutron star \cite{meta,third,Ell} in the RMF model, also in the density dependent RMF model\cite{ddrh}. 
The net effect of $K^-$ condensates in neutron star matter is to maintain charge neutrality replacing electrons and to soften the EoS 
resulting in the reduction of maximum mass of the neutron star \cite{meta,ddrh}, which was found to be within the observational limit. 
Also the threshold of (anti)kaon condensation is sensitive to antikaon optical potential and presence of charged hyperons pushes the 
threshold to higher densities. In a recent study both the approaches - density dependent couplings and higher order couplings in 
presence of (anti)kaon condensates have been compared \cite{gupta}. All the parameter sets were found to produce $2M_{solar}$ neutron 
stars without antikaon condensate and some with antikaon condensate, but hyperons were not included in that study.

In this paper, we investigate the possibility of antikaon condensation in beta equilibrated hyperon matter relevant to the 
dense interior of compact stars. Here we work with less to moderately attractive antikaon optical potential depth. 
We also use $\phi$-meson for hyperonic and kaonic interaction. Antikaon condensation in the presence of hyperon with additional $\phi$-meson
has been studied previously \cite{ddrh}, but not in the realistic density dependent framework. In this work we are interested to explore 
in a density dependent model whether this softening of EoS that arises due to both antikaon condensation and hyperon, can still produce 
a $2M_{solar}$ neutron star within the observational limit. The paper is organized as follows. In Section I, we briefly describe the
model to calculate the EoS. The parameters of the model are listed in Section III. Section IV is devoted to results and discussion. Finally 
we summarise in Section V.

\section{Formalism}
A phase transition from hadronic to antikaon condensed matter is considered here.
This phase transition could be either a first order or second order transition.
The hadronic phase is made of different species of the baryon octet along with
electrons and muons making a uniform background.
In the present approach, the model Lagrangian density (${\cal L} = {\cal L}_B + {\cal L}_l$) is of the form
\begin{eqnarray}
\label{eq_lag_b}
{\cal L}_B &=& \sum_{B= N, \Lambda, \Sigma, \Xi} \bar\psi_{B}\left(i\gamma_\mu{\partial^\mu} - m_B
+ g_{\sigma B} \sigma - g_{\omega B} \gamma_\mu \omega^\mu 
-  g_{\rho B} 
\gamma_\mu{\mbox{\boldmath $\tau$}}_B \cdot 
{\mbox{\boldmath $\rho$}}^\mu  \right)\psi_B\nonumber\\
&& + \frac{1}{2}\left( \partial_\mu \sigma\partial^\mu \sigma
- m_\sigma^2 \sigma^2\right)
-\frac{1}{4} \omega_{\mu\nu}\omega^{\mu\nu}\nonumber\\
&&+\frac{1}{2}m_\omega^2 \omega_\mu \omega^\mu
- \frac{1}{4}{\mbox {\boldmath $\rho$}}_{\mu\nu} \cdot
{\mbox {\boldmath $\rho$}}^{\mu\nu}
+ \frac{1}{2}m_\rho^2 {\mbox {\boldmath $\rho$}}_\mu \cdot
{\mbox {\boldmath $\rho$}}^\mu.
\end{eqnarray} 
Leptons are treated as non-interacting particles and described by the Lagrangian density
\begin{eqnarray}
{\cal L}_l &=& 
\sum_l \bar\psi_l\left(i\gamma_\mu {\partial^\mu} - m_l \right)\psi_l ~.
\end{eqnarray}
Here $\psi_l$ ($l \equiv {e,\mu}$) is lepton spinor whereas $\psi_B$ denotes 
the baryon octet. Baryons interact via the exchange of scalar $\sigma$, vector $\omega$,
$\rho$ mesons; ${\mbox{\boldmath $\tau_{B}$}}$ is the isospin operator. 
The field strength tensors for the vector mesons are given by
$ \omega^{\mu \nu}= \partial^ \mu \omega^ \nu-\partial^\nu \omega^ \mu$
and
${\boldsymbol \rho^{\mu \nu}}= \partial^ \mu {\boldsymbol \rho}^ \nu-\partial^\nu {\boldsymbol \rho}^ \mu $.
The $g_{\alpha B}(\hat n)$'s, where $\alpha=\sigma,~\omega$ and $\rho$
specify the coupling strength of the mesons with baryons and are vector density-dependent. The density operator 
$\hat {n}$ has the form, 
$\hat n$=$\sqrt{{\hat j}_\mu {\hat j}^{\mu}}$, where ${\hat j}_\mu = \bar 
\psi \gamma_\mu \psi$.
Also, the meson-baryon couplings 
become function of total baryon density $n$ i.e. $<g_{\alpha B}(\hat n)>=g_{\alpha B}(<\hat n>)=g_{\alpha B}(n)$ \cite{ddrh,typel09}.

The Lagrangian structure closely follows the formalism of Typel et al. 
\cite{typel05, typel09}.  The above model has been extended to accommodate 
the whole baryon octet. The interaction of hyperons with the nucleons is 
considered through meson exchange just like the nucleon-nucleon interaction. 
However, 
an additional vector meson $\phi$ and a scalar meson $\sigma^*$ are also included,
they are important for the the hyperon-hyperon interaction only 
\cite{Sch,Sch94}. Interaction among hyperons can be represented  
by the Lagrangian density
\begin{eqnarray} \label{eq_lag_y}
{\cal L}_{YY}&=&\sum_B \bar\psi_{B}\left(g_{\sigma^* B} \sigma^* 
 - g_{\phi B} \gamma_\mu \phi^\mu
\right)\psi_B \nonumber\\
&& + \frac{1}{2}\left( \partial_\mu \sigma^* \partial^\mu \sigma^*
- m_{\sigma^*}^2 {\sigma^*}^2\right) \nonumber\\
&& -\frac{1}{4} \phi_{\mu\nu}\phi^{\mu\nu}
+\frac{1}{2}m_\phi^2 \phi_\mu \phi^\mu~.
\end{eqnarray}
It has been reported that the attractive hyperon-hyperon interaction mediated
by $\sigma^*$ meson is very weak \cite{Sch}. We neglect the contribution of 
$\sigma^*$ meson in this calculation.

Using Euler-Lagrange relation the equations of motion for the meson and baryons
fields are easily derived from the total Lagrangian density (${\cal L} = {\cal L}_B + {\cal L}_l +{\cal L}_{YY}$).
The density dependence of the couplings while computing  
variation of $\cal L$ with respect to $\psi_B$ gives rise to
an additional term, which we denote by the rearrangement term \cite{ddrh, typel09}. The meson field equations are solved self-consistently
keeping into consideration the conditions for charge neutrality and 
baryon number conservation. We consider a static and isotropic matter 
in the ground state.
For such a static system, all space and time derivatives of
the fields vanish. Also, in the rest frame of the matter the space components of
 $\omega_{\mu}$, $\rho_{\mu}$  and $\phi_{\mu}$ vanish.
Furthermore, the third component of the isovector $\rho$ meson couples to baryons
because the expectation values of the sources for charged $\rho$ mesons in the equation of motion also vanish in the
ground state. It is to be noted $\phi$ mesons do not couple with nucleons i.e. $g_{\phi N}=0$. The meson field equations are solved 
in the mean-field approximation where the meson fields are replaced by their expectation values. 
The meson field equations are given by
\begin{eqnarray}
m_\sigma^2\sigma &=& 
\sum_B g_{\sigma B} n_B^{s}~,\\
m_\omega^2\omega_0 &=& \sum_B g_{\omega B} n_B,\\
m_\rho^2\rho_{03} &=& \sum_B g_{\rho B} \tau_{3B} n_B~,\\
m_\phi^2\phi_0 &=& \sum_B g_{\phi B} n_B~,
\end{eqnarray} 
The number density and scalar number density for the baryon B are given by
\begin{eqnarray}
n_B &=& <\bar \psi_B \gamma_0 \psi_B>=\frac{k_{F_{B}}^3}{3 \pi^2}~,\\
n^{s}_B &= & <\bar \psi_{B} \psi_B>= \frac{2J_B+1}{2\pi^2} \int_0^{k_{F_B}}
\frac{m_B^*}{(k^2 + m_B^{* 2})^{1/2}} k^2 \ dk \nonumber \\
&&=\frac{m_B^*}{2\pi^2}[k_{F_{B}}\sqrt{{k_{F_B}}^2+m_B^{*2}} - m_B^{*2}ln 
\frac {k_{F_{B}}+\sqrt{{k_{F_B}}^2+m_B^{*2}}}{m_B^*}]~.
\end{eqnarray} 
The Dirac equation for the spin $\frac 1 2$ particles is given by
\begin{equation}
[\gamma_{\mu}\left(i\partial^{\mu}-\Sigma_{B}\right)-m_B^*
]\psi_{B}=0.
\end{equation}
The effective baryon mass is defined as 
$m_B^*=m_B-g_{\sigma B}\sigma$, with $m_B$ as the vacuum rest mass of baryon B whereas
$\Sigma_{B}=\Sigma^{(0)}_{B}+\Sigma^{(r)}_B$ is the vector self-energy. 
The first term in the vector 
self-energy consists of the usual non-vanishing components of the vector mesons 
i.e. $\Sigma^{(0)}_{B}=g_{\omega B}\omega_0 + g_{\rho B} \tau_{3B} \rho_{03}+g_{\phi B} \phi_{0}~.$ 
while the second term is the rearrangement term, which arises due to the density-dependence of meson-baryon couplings \cite{ddrh}, 
assumes the form
\begin{equation}\label{eq_rear}
\Sigma^{(r)}_B=\sum_B[-g_{\sigma B}'  
\sigma n^{s}_B + g_{\omega B}' \omega_0 
n_B
+ g_{\rho B}'\tau_{3B} \rho_{03} n_B
+   g_{\phi  B}' \phi_0 n_B  ]~,
\end{equation}
where $g_{\alpha B}'=\frac {\partial g_{\alpha B}} {\partial \rho_B}$, $\alpha= \sigma,~ \omega, ~\rho, ~\phi$ and $\tau_{3B}$ is 
the isospin projection of $B=n, p, \Lambda, \Sigma^-,\Sigma^0, \Sigma^-,\Xi^-,\Xi^0$.
In the interior of neutron stars, the baryons and leptons are
in chemical equilibrium governed by the general equilibrium condition $\mu_i = b_i \mu_n - q_i \mu_e ~$, where $b_i$ is the baryon number 
and $q_i$, the charge of $i$th baryon and $\mu_n$ is the chemical potential of 
neutron and $\mu_e$ is that of electron. 
This condition determines the threshold of a particular hyperon. As the chemical potential of the neutron and electron 
becomes sufficiently large at high density and eventually the threshold of 
hyperons is reached, they are populated. 
The chemical potential for the baryon B is
$\mu_{B} = \sqrt{k_B^2+m_B^{*2} }+ g_{\omega B} \omega_0 + g_{\rho B} \tau_{3B}\rho_{03} + g_{\phi B}\phi_0+ \Sigma^{(r)}_B$. 
The term $g_{\phi B}\phi_0$ in $\mu_B$ is applicable for hyperons only.
The energy density due to baryons can be explicitly expressed as
\begin{eqnarray}\label{eq_ener_h}
{\varepsilon_B}  &=& \frac{1}{2}m_\sigma^2 \sigma^2
+ \frac{1}{2} m_\omega^2 \omega_0^2+ \frac{1}{2} m_\rho^2 \rho_{03}^2+ 
 \frac{1}{2} m_\phi^2 \phi_{0}^2+ \nonumber \\
&&  \sum_B \frac{2J_B+1}{2\pi^2} 
\int_0^{k_{F_B}} (k^2+m^{* 2}_B)^{1/2} k^2 \ dk
+ \sum_l \frac{1}{\pi^2} \int_0^{K_{F_l}} (k^2+m^2_l)^{1/2} k^2 \ dk. 
\end{eqnarray}
However, the expression for pressure in addition contains 
the rearrangement term ($\Sigma^{(r)}_B$) and is given by
\begin{eqnarray}
P_B &=& - \frac{1}{2}m_\sigma^2 \sigma^2
+ \frac{1}{2} m_\omega^2 \omega_0^2
+ \frac{1}{2} m_\phi^2 \phi_0^2
+ \frac{1}{2} m_\rho^2 \rho_{03}^2 +
\Sigma^{(r)}_B \sum_B n_B \nonumber \\
&& + \frac{1}{3}\sum_B \frac{2J_B+1}{2\pi^2}
\int_0^{k_{F_B}} \frac{k^4 \ dk}{(k^2+m^{* 2}_B)^{1/2}}
+ \frac{1}{3} \sum_l \frac{1}{\pi^2}
\int_0^{K_{F_l}} \frac{k^4 \ dk}{(k^2+m^2_l)^{1/2}} ~.
\end{eqnarray}
The pressure ($P_B$) is related to the energy density ($\varepsilon_B$)
in this phase through the Gibbs-Duhem relation
\begin{equation}
P_B=\sum_i \mu_i n_i -\varepsilon_B~.
\end{equation}
The rearrangement
term does not contribute to the energy density explicitly, whereas it occurs in
the pressure through baryon chemical potentials.
It is the rearrangement term that accounts for the energy-momentum conservation
and thermodynamic consistency of the system \cite{ddrh}.
Similarly, we calculate number densities, energy densities and pressures of
electrons and  muons.

Next we discuss the antikaon condensed phase composed of all the species
of the baryon octet, the antikaon isospin doublet with electron and muons in 
the background. The baryon-baryon interaction in the antikaon condensed phase 
is described by the Lagrangian density of Eq. (\ref{eq_lag_b}).
We choose the antikaon-baryon interaction on the same footing as the
baryon-baryon interaction.
The Lagrangian density for (anti)kaons in the minimal coupling scheme is given 
by \cite{Pal,meta,third,Gle99}
\begin{equation}
{\cal L}_K = D^*_\mu{\bar K} D^\mu K - m_K^{* 2} {\bar K} K ~,
\end{equation}
where the covariant derivative is
$D_\mu = \partial_\mu + ig_{\omega K}{\omega_\mu}
+ i g_{\rho K} {\boldsymbol \tau}_K \cdot {\boldsymbol \rho}_\mu + i g_{\phi K} {\phi_\mu}$ and the effective mass of (anti)kaons is given by
$m_K^* = m_K - g_{\sigma K} \sigma $
where $m_K$ is the bare kaon mass.
The isospin doublet for kaons
is denoted by $K\equiv (K^+, K^0)$ and that for antikaons is
$\bar K\equiv (K^-, \bar K^0)$.
For s-wave (${\bf p}=0$) condensation, the in-medium energies of
$\bar K\equiv (K^-, \bar K^0)$ 
are given by
\begin{equation} \label{ch1.kom}
\omega_{K^-,\: \bar K^0} = m_K^* - g_{\omega K} \omega_0 - g_{\phi K} \phi_0
\mp  g_{\rho K} \rho_{03}.
\end{equation}
It is to be noted that for $s$-wave ({\bf k}=0) ${\bar K}$ condensation at T=0, 
the scalar and vector densities of antikaons are same and those are given by 
\cite{Gle99} 
\begin{equation}
n_{K^-,\: \bar K^0}=2\left( \omega_{K^-, \bar K^0} + g_{\omega K} \omega_0
+ g_{\phi K} \phi_0 \pm  g_{\rho K} \rho_{03} \right) {\bar K} K~.
\end{equation} 
The requirement of chemical equilibrium fixes the
onset condition of antikaon condensations in neutron star matter.
\begin{eqnarray}
\mu_n - \mu_p &=& \mu_{K^-} = \mu_e ~, \\
\mu_{\bar K^0} &=& 0 ~,
\end{eqnarray}
where $\mu_{K^-}$ and $\mu_{\bar K^0}$ are respectively the chemical
potentials of $K^-$ and $\bar K^0$.
In the mean field approximation, the meson field equations in the presence of 
antikaon condensates are given by
\begin{eqnarray}
m_\sigma^2\sigma &=& 
\sum_B g_{\sigma B} n_B^{s}
+ g_{\sigma K} \sum_{\bar K} n_{\bar K} ~,\\
m_\omega^2\omega_0 &=& \sum_B g_{\omega B} n_B
- g_{\omega K} \sum_{\bar K} n_{\bar K} ~,\\
m_\rho^2\rho_{03} &=& \sum_B g_{\rho B} \tau_{3B} n_B
+ g_{\rho K} \sum_{\bar K} \tau_{3 {\bar K}} n_{\bar K}~,\\
m_\phi^2\phi_0 &=& \sum_B g_{\phi B} n_B
- g_{\phi K} \sum_{\bar K} n_{\bar K} ~,
\end{eqnarray}
Antikaon condensates do not directly contribute to the pressure so 
it is due to baryons and
leptons only. However, the presence of additional term due to (anti)kaons 
in the meson field equations change the fields. Also $K^-$ mesons modify 
the charge neutrality condition. 
Thus the values of 
rearrangement term, pressure etc. are changed when the (anti)kaons appear.
The energy density of (anti)kaons is given by $\epsilon_{\bar K} = m^*_K \left( n_{K^-} + n_{\bar K^0}\right)$. 
The total energy density  has contribution from the baryons, antikaons and 
leptons $\epsilon = \epsilon_B + \epsilon_{\bar K} + \epsilon_l$.

\section{Model Parameters}
The nucleon-meson density-dependent couplings are determined following the 
prescription of Typel et. al \cite{typel05,typel09}. 
The functional dependence of the couplings on density was first introduced in 
\cite{typel99} and is described as 
\begin{equation}
g_{\alpha B}(n_b)=g_{\alpha B}(n_{0}) f_{\alpha}(x),
\end{equation}
where $n_b$ is the total baryon density defined as, $n_b = \sum_B n_B$ , $x=n_b/n_0$, and 
$f_{\alpha}(x)=a_{\alpha}\frac{1+b_{\alpha}(x+d_{\alpha})^2}{1+c_{\alpha}
(x+d_{\alpha})^2}$ is taken for $\alpha=\omega$, $\sigma$. the number of parameters are reduced by constraining the functions 
as $f_{\sigma}(1)=f_{\omega}(1)=1$, $f_{\sigma}'(0)=f_{\omega}'(0)=0$ and $f_{\sigma}(1)=f_{\omega}(1)=1$
, $f_{\sigma}''(1)=f_{\omega}''(1)$ \cite{typel05}.
The $\boldsymbol \rho_{\mu}$ coupling decreases at higher densities, therefore,
an exponential density-dependence is assumed  for the isovector meson ${\rho}$ i.e. 
$f_{\alpha}(x) = \exp[{-a_{\alpha} (x-1)}]$ \cite{typel99}.
These functional dependence is now widely used \cite{Nik02,Lal05,Armen}. 
The saturation 
density, the mass of $\sigma$ meson,  the couplings $g_{\alpha B}(n_{0})$ and 
the coefficients $a_{\alpha}$,$b_{\alpha}$,$c_{\alpha}$,$d_{\alpha}$ are found by fitting the finite nuclei properties 
\cite{typel05,typel09} and are tabulated in Table \ref{tab1}.
The fit gives the saturation density $n_{0}=0.149065fm^{-3}$, 
binding energy per nucleon as $-16.02$MeV and incompressibility $K=242.7$MeV.
The masses of neutron, proton, $\omega$ and $\rho$ mesons are 939.56536, 
938.27203, 783 and 763 MeV respectively (See Table II of 
Ref \cite{typel09}). 
\begin{table}

\caption{Parameters of the meson-nucleon couplings in DD2 model}
\label{tab1}
\begin{tabular}{|c|ccccc|} 

\hline
meson $\alpha$& $g_{\alpha B}$ &$a_{\alpha}$&$b_{\alpha}$&$c_{\alpha}$&$d_{\alpha}$\\
\hline
$\omega$&13.342362&1.369718&0.496475&0.817753&0.638452\\
$\sigma$&10.686681&1.357630&0.634442&1.005358&0.575810\\
$\rho$&3.626940&0.518903&&&\\
\hline
\end{tabular}
\end{table}

Next we determine the hyperon-meson couplings.
In the absence of density-dependent Dirac-Bruekner 
calculation for hyperon couplings, we use scaling factors \cite{Sch} and 
nucleon-meson couplings of Table \ref{tab1} to determine the 
hyperon-meson couplings. 
The vector coupling constants for hyperons are determined from the SU(6)
symmetry \cite{Sch} as,
\begin{eqnarray}
\frac{1}{2}g_{\omega \Sigma} = g_{\omega \Xi} =
\frac{1}{3} g_{\omega N},\nonumber\\
\frac{1}{2}g_{\rho \Sigma} = g_{\rho \Xi} = g_{\rho N}{\rm ;}~~~
g_{\rho \Lambda} = 0, \nonumber\\
2 g_{\phi \Lambda} =  g_{\phi \Xi} =
-\frac{2\sqrt{2}}{3} g_{\omega N} ~.
\end{eqnarray}
The scalar meson ($\sigma$) coupling to hyperons is obtained from the
potential depth of a hyperon (Y) in the saturated nuclear matter
\begin{equation}
U_Y^N(n_0) = - g_{\sigma Y} {\sigma} + g_{\omega Y} {\omega_0} +\Sigma^{(r)}_N, 
\end{equation}
where $\Sigma^{(r)}_N$ involves only the contributions of nucleons.
The analysis of energy levels in $\Lambda$-hypernuclei suggests a potential well depth 
of $\Lambda$ in symmetric matter $U_{\Lambda}^N(n_0)=-30$ MeV \cite{ch1.Chr,ch1.Dov}.
On the other hand, recent analysis of a few $\Xi$-hypernuclei events predict a $\Xi$ well depth of
$U_{\Xi}^N(n_0)=-18$ MeV \cite{ch1.Fuk,ch1.Kha}. However, $\Sigma$ hyperons are ruled out because of the repulsive $\Sigma-$potential
depth in nuclear matter. The  particular choice of hyperon-nucleon potential does not change the maximum mass of neutron 
stars \cite{weiss1}. We use these values and find the scaling factor as  $R_{\sigma \Lambda}=\frac{g_{\sigma \Lambda}}{g_{\sigma N}}=0.62008$ and $R_{\sigma \Xi}=\frac{g_{\sigma \Xi}}{g_{\sigma N}}=0.32097$. 
Finally we compute the meson-anti(kaon) couplings on the same footing as 
that of meson-hyperon couplings. However, we do not consider any 
density-dependence here. Coupling constants of $\omega$ and $\rho$ mesons with 
kaons are obtained from the quark model and isospin counting rule 
\cite{third,Gle99} and the coupling constant of $\phi$ mesons with kaons  
is given by the SU(3) relations and the value of $g_{\pi\pi\rho}$ \cite{Sch},
\begin{equation}
g_{\omega K} = \frac{1}{3} g_{\omega N} {\rm ;}~~~~~ 
g_{\rho K} = g_{\rho N} ~~~~~ {\rm and} ~~~~~ \sqrt{2} ~g_{\phi K} = 6.04.
\end{equation}
The scalar coupling constant ($g_{\sigma K}$) is obtained from the real part of the
$K^-$ optical potential at the normal nuclear matter density \cite{Sch,meta,third,ddrh}
\begin{equation}
U_{\bar K} \left(n_0\right) = - g_{\sigma K}\sigma - g_{\omega K}\omega_0 ~
+ \Sigma^{(r)}_N ~.
\end{equation}
The study of kaon atoms clearly suggests an attractive (anti)kaon nucleon optical potential. However,
there is controversy about how deep the potential is, whether the (anti)kaon 
optical potential is extremely deep, as it is preferred by the 
phenomenological fits to kaonic atoms data, or shallow, as it comes out from unitary chiral model calculations. Different experiments 
also suggest a range of values for $U_{\bar K}$ from $-50$ to $-200$MeV and do not come to any definite consensus\cite{ramos}. 
We chose a set of values of $U_{\bar K}$ from -60 to -140 MeV.
The coupling constants for kaons with $\sigma$-meson, $g_{\sigma K}$
at the saturation density for these values of $U_{\bar K}$ 
for DD2 model is listed in Table
\ref{tab2}.

\begin{table}

\caption{Parameters of the scalar $\sigma$ meson -(anti)kaon couplings in DD2 model}
\label{tab2}
\begin{tabular}{|c|ccccc|} 

\hline
$U_{\bar K}$ (MeV)&-60&-80&-100&-120&-140\\
\hline
$g_{\sigma \bar K}$&-1.24609&-0.72583&-0.20557&0.31469&0.83495\\
\hline
\end{tabular}
\end{table}

\section{Results}

We report our results calculated using the DD2 model. We begin with the 
composition of the 
star in the presence of different exotic particles. As the neutron chemical 
potential and the Fermi level of nucleons become sufficiently large at high 
density, different exotic particles could be populated in the core of the star. 
First we consider antikaon condensates ($K^-, \bar K^0$) 
in the nucleon-only system consisting of proton, neutron, electron and muon. 
For $U_{\bar K}(n_0)=-60$MeV, $K^-$ appears at $4.11n_0$ in the nucleon-only 
matter. The threshold density of $K^-$ condensation decreases as the antikaon potential in nuclear matter becomes more attractive.
We note that the threshold density of $\bar K$ condensation shifts towards lower density as the strength of 
$|U_{\bar K}(n_0)|$ increases. Also, it is observed that $K^-$ condensates 
populate before $\bar K^0$ condensate appears. It is always energetically 
favorable to 
populate the condensates of negatively charged kaons, that takes care 
of the charge 
neutrality but being condensates, do not add to the pressure unlike the 
leptons. 
The threshold densities of the $K^-(\bar K^0)$ in $\beta$-equilibrated matter 
with different compositions
are listed in Table \ref {tab3}, the values corresponding to $\bar K^0$ 
condensates are given in the parentheses. 

Next, we consider $\Lambda$ and $\Xi^-$, $\Xi^0$ apart from the nucleons.
At low density, the system consists of only nucleon and leptons until strange baryons appear beyond twice the normal matter density.
$\Lambda$ hyperons are the first to appear at $2.22n_0$, followed by  $\Xi^-$ 
at $2.44n_0$ and 
finally $\Xi^0$ sets in at $7.93n_0$. If we allow the (anti)kaons
in addition to $\Lambda$ hyperons, $K^-$ appears at $3.07n_0$ and $6.54n_0$
at $U_{\bar K}=-140$MeV  and $-60$ MeV, respectively. However, $\bar K^0$ 
appears only at higher 
density and for a deeper potential depth ($|U_{\bar K}| \geq  120$ MeV). 
The presence of hyperons delays the onset of $\bar K$ condensation 
to higher density as evident from Table \ref{tab3}. Moreover, 
negatively 
charged hyperons diminish the electron chemical 
potential delaying the onset of $K^-$ condensation.  

\begin{figure}
\includegraphics[width=0.9\textwidth]{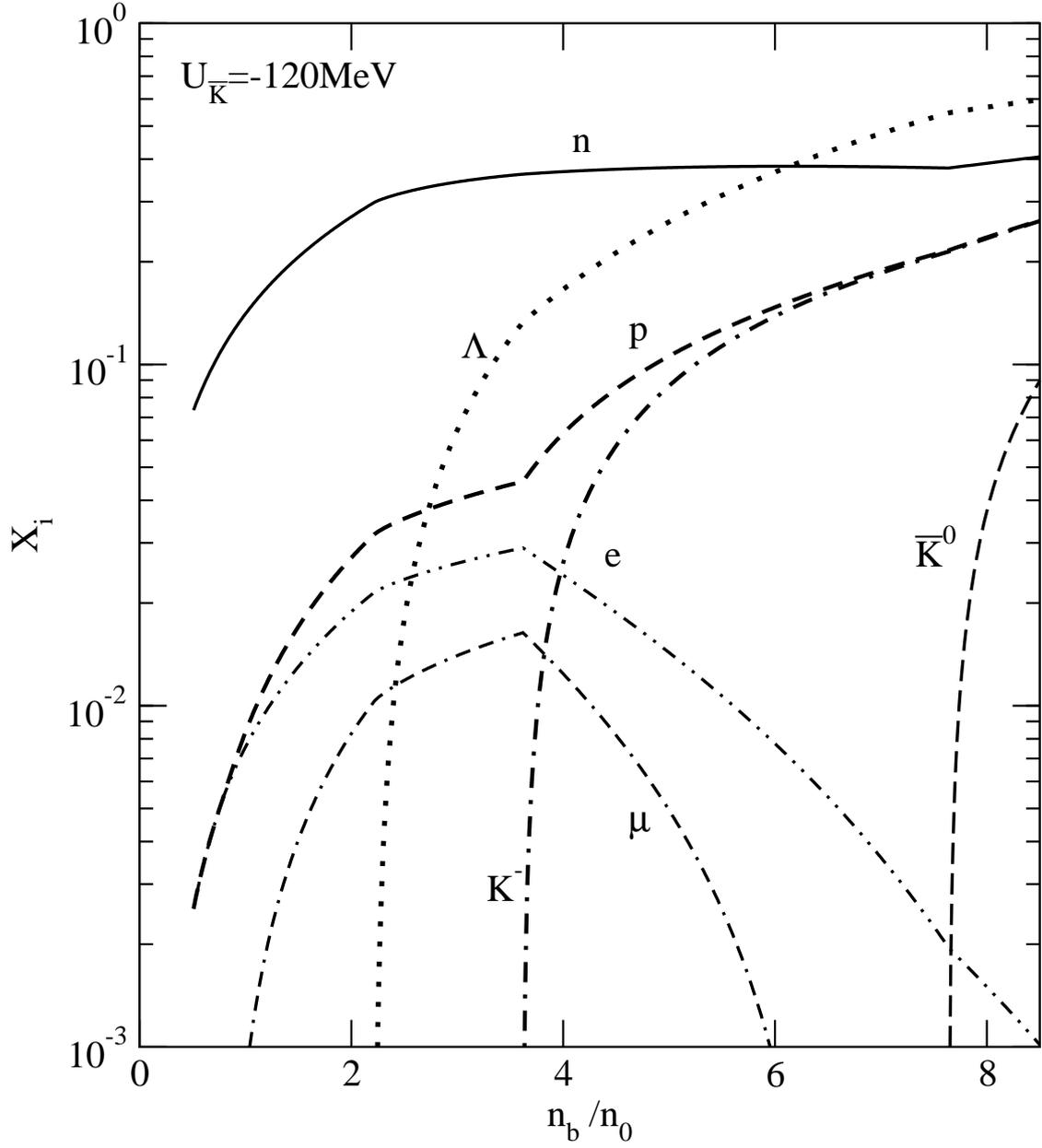}
\caption{Fraction of various particles in $\beta$-equlibrated n, p, $\Lambda$ and lepton matter including $K^-$ and $\bar K^0$ condensates
for $U_{\bar K}(n_0)=-120$ MeV as a function of normalised baryon density.}
\label{part_npkl}
\end{figure}

\begin{figure}
\includegraphics[width=0.9\textwidth]{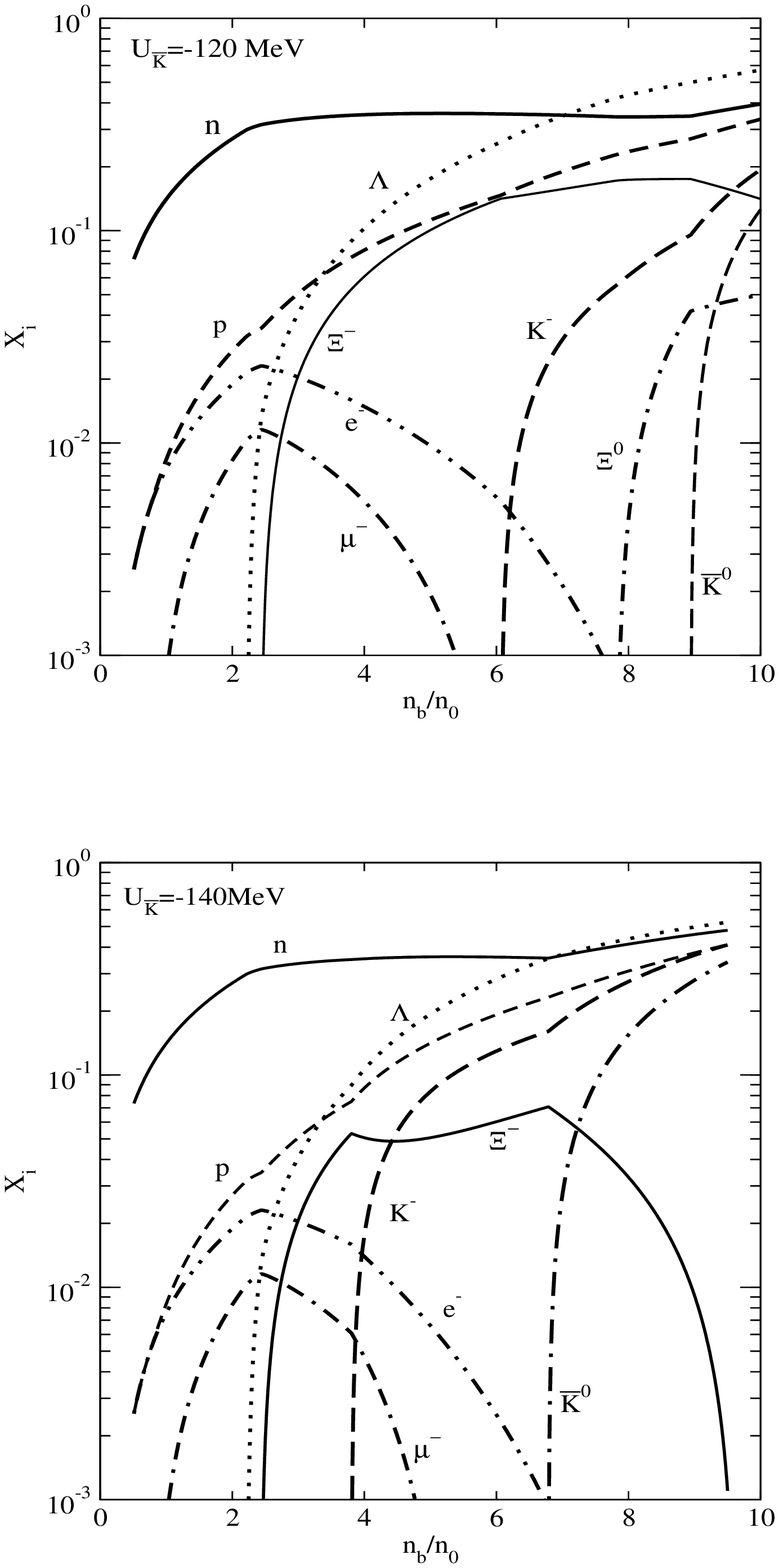}
\caption{Fraction of various particles in $\beta$-equlibrated n, p, $\Lambda$, $\Xi^-$, $\Xi^0$ and lepton matter including $K^-$ and $\bar K^0$ condensates
for $U_{\bar K}(n_0)=-120$ MeV and $-140$ MeV as a function of normalised baryon density.}
\label{part_npklc}
\end{figure}

In Fig. \ref{part_npkl} we compare the particle fractions for a particular 
value of $U_{\bar K}=-120$MeV. 
Before the onset of exotic particles, the charge neutrality is maintained among protons, electrons and muons.
We see that $\Lambda$ hyperons appear at 2.22$n_0$ and its density rises fast at the 
cost of neutrons.  We notice that the onset of $K^-$ condensates takes care of 
the charge neutrality of the system as soon as it appears at $3.63n_0$ and 
leptons are depleted.  This behaviour is quite expected, as  $K^-$ mesons, 
being bosons, condense in the lowest energy state and are 
therefore energetically favorable to maintain the charge neutrality of the 
system.  Another notable fact is the rise of proton fraction 
as soon as the $K^-$ condensate takes care of the negative charge neutrality; 
leads to an almost iso-spin symmetric matter at higher density.
In case $\Xi^-$ is also present, both the (anti)kaons condense only at higher 
density and for $|U_{\bar K}| \geq  120$ MeV as is noticed in Fig. 
\ref{part_npklc}.  The early 
onset of $\Xi^-$ hyperons does not allow $\bar K$ to appear in the system for lower values of $U_{\bar K}$. We 
see the competition of all the exotic particles in 
Fig. \ref{part_npklc} for $U_{\bar K}=-120~ {\rm and}~-140$MeV. Though the 
onset of $\Xi^-$ delays the appearance of antikaon condensates, with 
stronger $U_{\bar K}=-140$MeV, $K^-$ suppresses $\Xi^-$ and even manages to replace it completely at higher density.

In Fig. \ref{eos_npk} pressure (P) is plotted against energy density ($\epsilon$) 
for system consisting of nucleons  and (anti)kaons 
for different $U_{\bar K}$. The solid line corresponds to the nucleon-only matter whereas the other lines 
correspond to the matter including $K^-$ and $\bar K^0$ 
condensates for antikaon optical potentials $U_{\bar K}(n_0)$ = -60 to -140 MeV.
The EoS is softened as soon as the $K^-$ and $\bar K^0$ appear, the effect 
being more pronounced for a deeper $U_{\bar K}$. The EoS with 
$U_{\bar K}=-140$MeV is the softest. 
The kinks in the EoS at mid energy densities ($426.5~ {\rm to}~ 693.0 MeV fm^{-3}$) 
correspond to the $K^-$ onset and those at higher densities 
($872.1~ {\rm to}~ 1492.6 MeV fm^{-3}$) mark the $\bar K^0$ condensation. 

\begin{figure}
\includegraphics[width=0.9\textwidth]{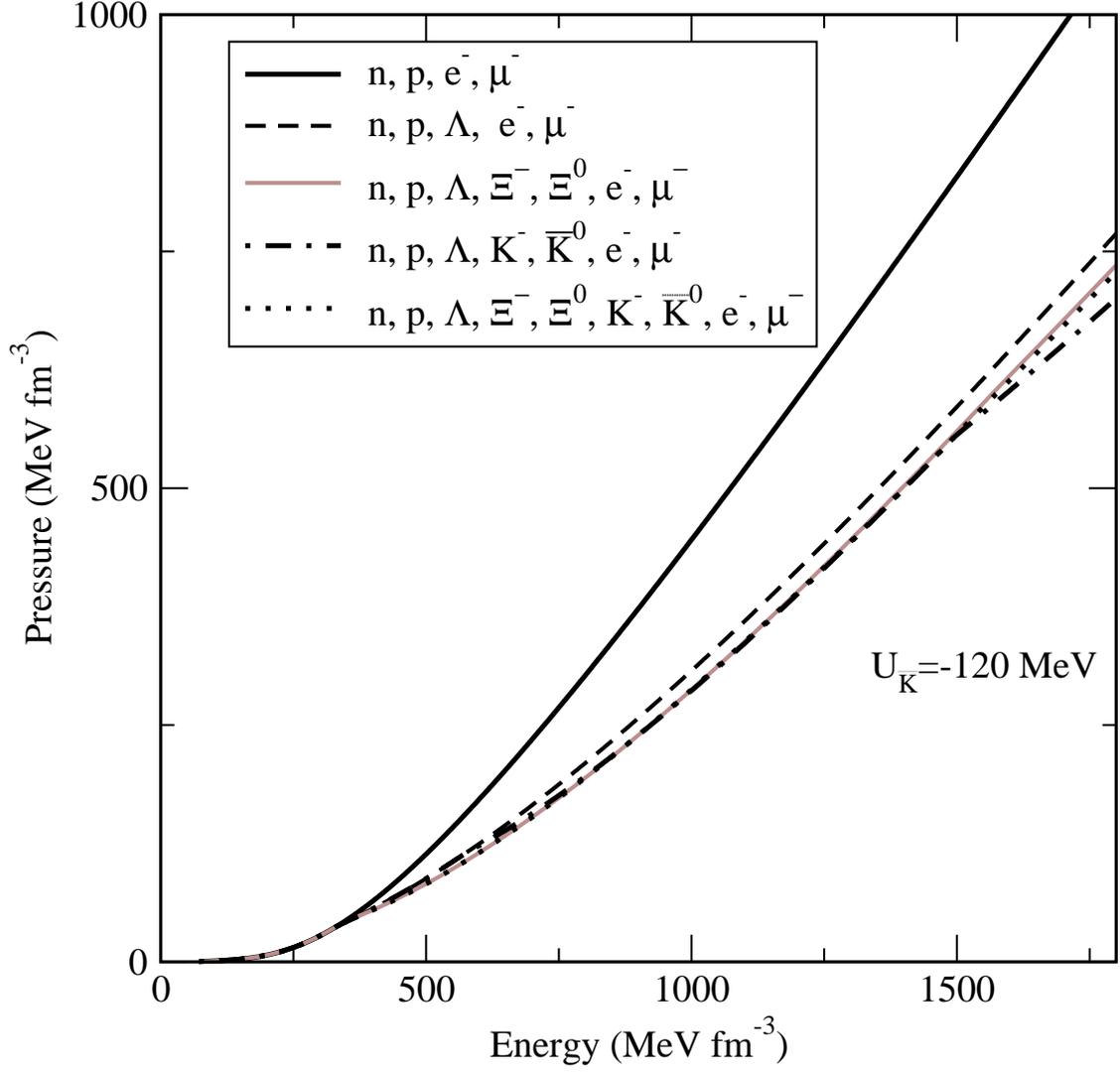}
\caption{The equation of state (EoS), pressure (P) vs energy density ($\epsilon$) for various particle combination of n, p, $\Lambda$, $\Xi^-$, $\Xi^0$ and 
lepton in $\beta$-equilibrated matter including $K^-$ and $\bar K^0$ condensates with $U_{\bar K}(n_0)$=-120 MeV.}
\label{eos_npklc}

\end{figure}

Similarly we draw the EoS in the presence of additional hyperons in Fig. \ref{eos_npklc}. 
With the appearance of $\Lambda$ hyperons at $330{\rm MeV} fm^-3$, the slope of the EoS deviates from the nucleon one. The EoS is 
further softened at the onset of  $\Xi^-$. However, the EoS considering 
all the exotic particles is not the softest one here.  
We have seen that hyperons delay (anti)kaons to higher density.  
This explains the relative stiffness of the EoS at higher density in the presence of $\Xi$ along 
with other particles. In the figure we only draw the (anti)kaon EoS 
corresponding to $U_{\bar K}(n_0)=-120$MeV.  

\begin{table}

\caption{Threshold density (in units of $n_0$) of the $K^-$ ($\bar K^0$) condensates in the DD2 model. (-) denotes no-show of them.}
\label{tab3}
\begin{tabular}{|c|lllll|} 

\hline
$U_{\bar K}$ (MeV)&-60&-80&-100&-120&-140\\
\hline 
np$K^-\bar K^0$&4.11(7.16)&3.74(6.62)&3.40(6.07)&3.08(5.54)&2.79(5.00) \\
np$\Lambda K^-\bar K^0$&6.54(-)&5.30(-)&4.35(-)&3.63(7.65)&3.07(6.40)\\
np$\Lambda \Xi^- \Xi^0 K^-\bar K^0$&-(-)&-(-)&-(-)&6.07(8.95)&3.81(6.79)\\
\hline
\end{tabular}
\end{table}

\begin{figure}
\includegraphics[width=0.9\textwidth]{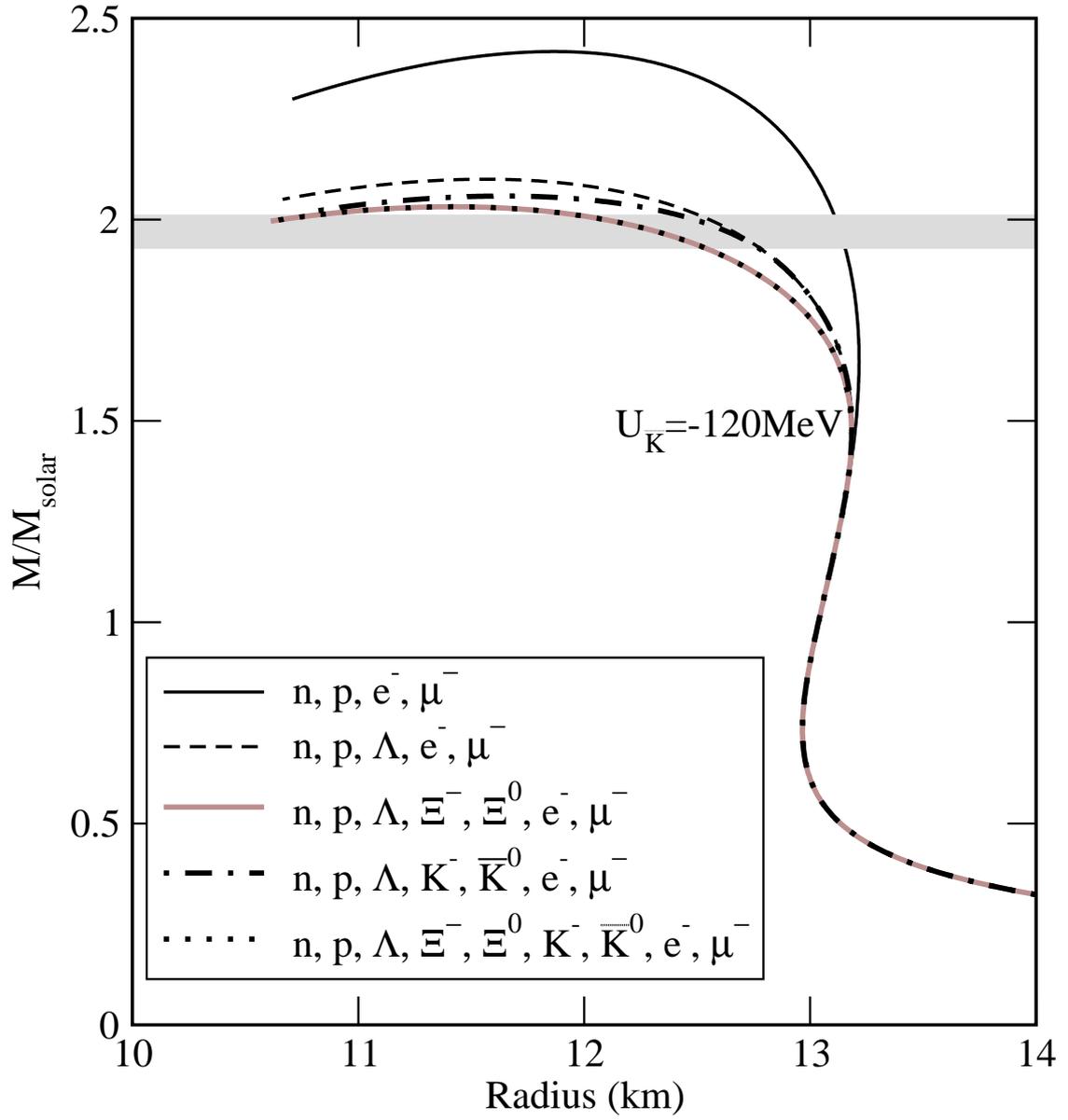}
\caption{The neutron star mass sequences are plotted with radius for the equations of state of Fig. \ref{eos_npklc}. 
The gray band specifies the observational limits.}
\label{tov_npklc}

\end{figure}

We solve the Tolman-Oppenheimer-Volkoff (TOV) equations for spherically 
symmetric, static compact stars and show our result in Figs. \ref{tov_npk}, \ref{tov_npklc} corresponding to the equations of state 
of Fig \ref{eos_npk} and \ref{eos_npklc}, respectively. For low density ($n < 0.001 fm^{-3}$) crust, 
we used the EoS of Baym, Pethick and Sutherland \cite{baym}. The set of maximum mass of the
nucleons-only and hyperon stars and their corresponding central densities and radii corresponding to EoS of Fig. \ref{eos_npklc}, 
are listed in Table \ref{tab4}. The gray band in both figures marks the observational limits of Refs. \cite{demo,anton}.
We notice that in all the cases the values of the maximum mass lie well above 
the benchmark $2.0~M_{solar}$, the radii being within the range of 11.42 to 
11.87 km. Radii decreases with additional exotic degrees of freedom. Softer 
the EoS, less mass it can support against gravity and more compact is the star.
The maximum mass of a nucleon-only star is $2.417M_{solar}$, with the inclusion of $\Lambda$ and $\Xi$ hyperons this reduces to $2.1M_{solar}$ 
and $2.032M_{solar}$ respectively. It is noted that the core contains 
$\Lambda$ and $\Xi^-$, but no $\Xi^0$ and is denser compared to the nucleon-only 
case.

Table \ref{tab5} enlists the values of maximum mass and its
corresponding central energy density and radius for the
hyperons and (anti)kaons EoS with different values of optical potential. 
When we consider (anti)kaons in addition to the nucleons, they are found to 
reduce 
the maximum mass of  the star for all $U_{\bar K}$, but the central density 
does not increase until it has got $\bar K^0$, 
which happens only  $|U_{\bar K}| \geq 120$ MeV. 
In the presence of $\Lambda$ hyperons, for $U_{\bar K}$ as low as -60 MeV, 
antikaons do not have any effect on the maximum mass, as $K^-$ condensate 
appears at $6.54n_0$, that is beyond the central density and $ \bar K^0$ does 
not appear at all. The effect of $K^-$ condensates is pronounced from $|U_{\bar K}| = 80$ MeV, where the core contains considerable fraction of $K^-$,  
but still no $\bar K^0$ condensates. Both the (anti)kaons appear only at 
$|U_{\bar K}| \geq 120$ MeV and reduce the maximum mass. 

Next we discuss the scenario
when our system contains $\Xi$'s in addition to nucleons, $\Lambda$ and $\bar K$. 
Though $\bar K$ appears for $|U_{\bar K}| \geq 120$ MeV, the maximum mass 
is reduced for $U_{\bar K}=-140$ MeV only. As it is evident from Fig. \ref{part_npklc}, the core (density $6.65n_0$) contains only $2\% ~{\rm and} ~15.5\%$ of 
$~K^-$ condensate for the two cases respectively whereas $\bar K^0 $ does not populate the core at all. So only $K^-$ condensate plays effective role in 
reducing the maximum mass of the star, that also for optical potential deeper 
than $-120$ MeV.

\begin{table}

\caption{Maximum mass, central density and radius of nucleons only as well as hyperon compact stars in the DD2 model. Maximum mass is in $M_{solar}$, central 
density with respect to the saturation density $n_0$, radius in km.}
\label{tab4}
\begin{tabular}{|clll|} 

\hline

&$M (M_{solar})$&$n_c~(n_0)$&R (Km)\\
np&2.417&5.71&11.87\\
np$\Lambda $&2.10&6.40&11.57\\
np$\Lambda \Xi$&2.032&6.66&11.42\\
\hline
\end{tabular}
\end{table}
 
\begin{table}

\caption{Maximum mass, central density and radius of compact stars with 
nucleons, hyperons and (anti)kaons for different values of
optical potential depth in the DD2 model. Maximum mass is in $M_{solar}$, central density in $n_0$, radius in km and $U_{\bar K}$ in MeV.}
\label{tab5}
\begin{tabular} {|c|lll|lll|lll|lll|lll|} 

\hline
$U_{\bar K}$&\multicolumn{3}{c|}{-60}&\multicolumn{3}{c|}{-80}&\multicolumn{3}{c|}{-100}&\multicolumn{3}{c|}{-120}&\multicolumn{3}{c|}{-140}\\
\hline
&M&$n_c$&R&M&$n_c$&R&M&$n_c$&R&M&$n_c$&R&M&$n_c$&R\\
\hline
np$K^- \bar K^0$&2.376&5.54&12.15&2.343&5.53&12.18&2.299&5.6&12.14&2.242&5.78&12.05&2.164&5.91&12.01\\
np$\Lambda K^- \bar K^0$&2.10&6.4&11.57&2.098&6.35&11.62&2.085&6.29&11.68&2.058&6.36&11.64&2.02&6.63&11.48\\
np$\Lambda \Xi^-\Xi^0 K^- \bar K^0$&2.032&6.66&11.42&2.032&6.66&11.42&2.032&6.66&11.42&2.032&6.65&11.43&2.016&6.67&11.4\\
\hline
\end{tabular}
\end{table}

\section{Summary}
We study  the equation of state and compositions of hyperons and antikaon 
condensates in neutron star matter within the framework of 
relativistic field theoretical  model with density-dependent couplings.  The density dependence of nucleon-meson couplings are 
determined following the DD2 model of Typel {\it et. al} \cite{typel05,typel09}. The density dependent meson-hyperon vertices are obtained 
from the density dependent meson-nucleon couplings using hypernuclei data \cite{Sch}, scaling law \cite{Kei} and SU(6) symmetry. The scalar meson 
coupling to $\Lambda$ and $\Xi$ hyperons are fitted to the potential depth of respective hyperons in saturated nuclear matter, which is 
available from experiments. A repulsive interaction between the hyperons are mediated by the exchange of $\phi(1020)$ mesons. The 
couplings of antikaon-nucleon interactions are obtained in the similar manner. However, they are not density-dependent.

The abundance of all the  particles considered here matches with the results of other models. In all the cases,  $\Lambda$ hyperons get 
into the system first, followed by the negatively charged $\Xi^-$ hyperons. The antikaon 
condensates also populate the nuclear matter at reasonably low densities for a deeper optical potential. However, in hyperon-rich matter
their appearance is delayed until higher densities. 
Also, the negatively charged hyperons diminish the electron chemical potential delaying the onset of $K^-$ condensation. All these 
findings are consistent with earlier results.   

Neutron star masses have been precisely measured for some binary pulsars. Until very recently, the largest precisely measured 
NS mass is $1.97 \pm 0.04 M_{solar}$ for PSR J1614−2230, and $2.01 \pm 0.04 M_{solar}$ for PSR J0348+0432.  We observe that the 
strangeness degrees of freedom softens the nuclear EoS that results into the reduction of neutron star maximum mass. Most of the existing 
models conflicts with the observation of such high pulsar masses. However, in 
all the cases we find the maximum mass within the constraint of observational 
limits. So we conclude that exotic EoS can not be ruled out by the observation 
of a $2M_{solar}$ compact star. In the framework of the DD2 model, there is a 
scope for accommodating strange hyperons and antikaon condensates within the observational limits of neutron star mass. This model can be 
exploited to develope a new EoS table involving antikaon condensates for core-collapse supernova explosions and neutron stars for a wide range of density, 
temperature and proton fraction.

As a final remark, we briefly mention the finite temperature effect on the 
hyperon EoS and maximum mass of the neutron stars. We notice a non-zero  temperature does not make much difference in the EoS and
maximum mass. But in the presence of $\Xi$ hyperons, the EoS differs slightly at finite temperature compared 
to the T=0 case. This is due to the late appearance of $\Xi^{-}$ and suppression of $\Xi^{0}$ in the former case. This difference is
found to have small effect on the mass-radius relation in both the cases. The maximum mass and 
corresponding radius in the presence of $n,~ p,~ \Lambda,~ \Xi^-, ~\Xi^0$  is
found to vary from  $2.032M_{solar}(11.42km)$ at T=0 to 
$2.108M_{solar}(11.72km)$ 
at T=15 MeV respectively. However, the transport properties of hot and $\beta$-equilibrated matter 
in neutron and proto-neutron stars might be affected, which on the other hand may have important implications for the thermal nucleation of droplets of antikaon condensed matter. We would report on the critical temperature of antikaon condensates in future.

\section{Acknowledgement}
We would like to thank Prof. D. Bandyopadhyay for useful discussions and remarks.

\newpage 

\begin{figure}
\includegraphics[width=0.9\textwidth]{eos_np_k.eps}
\caption{The equation of state (EoS), pressure (P) vs energy density ($\epsilon$). 
The full line is for n, p, and lepton matter whereas others are with 
additional $K^-$ and $\bar K^0$ condensates calculated with $U_{\bar K}(n_0)$= -60,-80,-100,-120 and -140 MeV. Deeper $U_{\bar K}$ corresponds to softer EoS.}
\label{eos_npk}
\end{figure}

\begin{figure}
\includegraphics[width=0.9\textwidth]{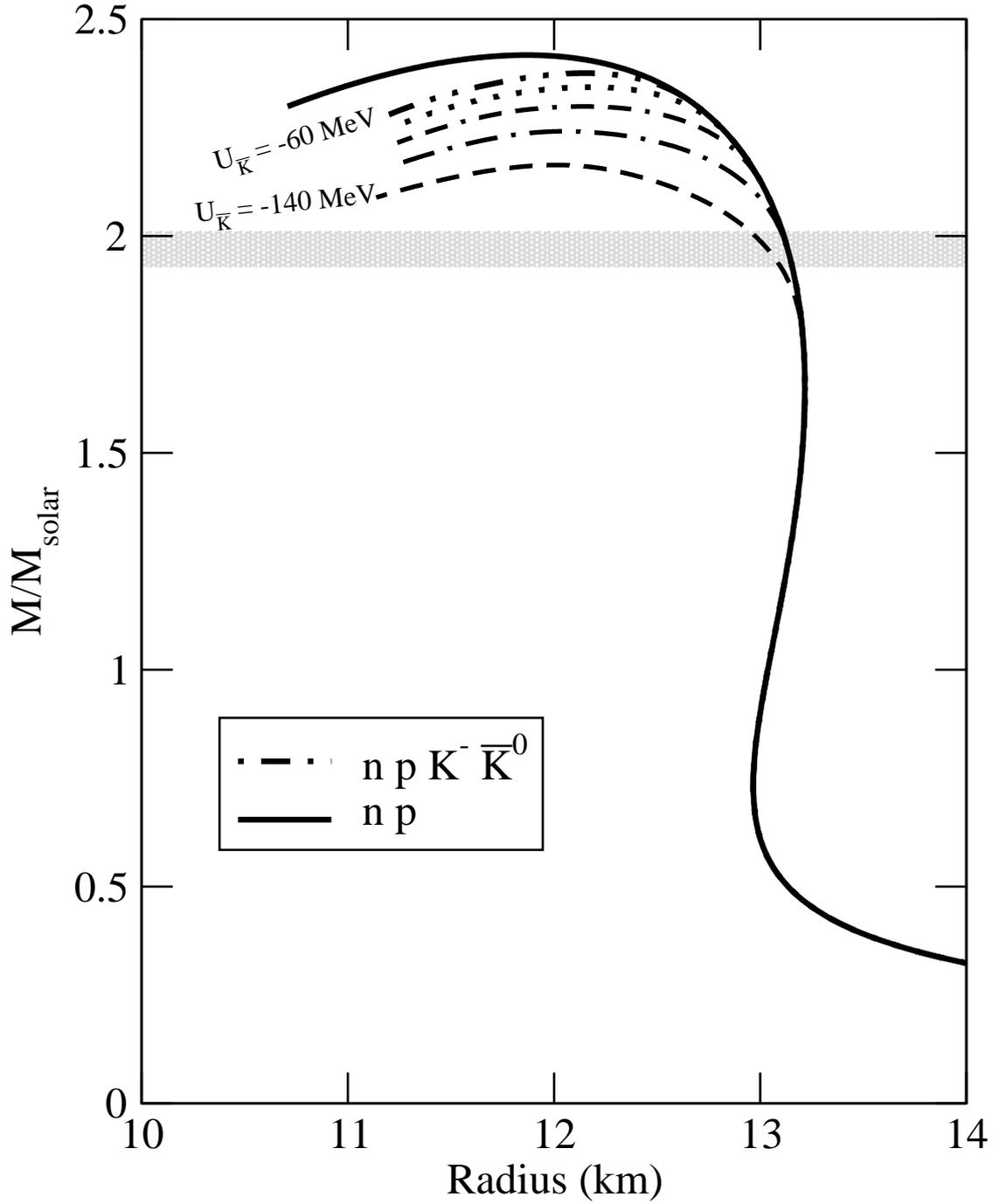}
\caption{The neutron star mass sequences are plotted with radius for the 
equations of state of Fig. \ref{eos_npk}. 
The full line is for n, p, and lepton matter whereas others are with 
additional $K^-$ and $\bar K^0$ condensates calculated with $U_{\bar K}(n_0)$= -60,-80,-100,-120 and -140 MeV. Deeper $U_{\bar K}$ corresponds to lower line.
The gray band specifies the observational limits.}
\label{tov_npk}

\end{figure}

\end{document}